\documentclass{aastex631}

\usepackage{amsthm,amsmath,amssymb}
\usepackage{lipsum}
\usepackage{float}
\usepackage{soul}

\hypersetup{linkcolor=red,citecolor=blue,filecolor=cyan,urlcolor=magenta}

\begin{document}

\title{BSN: Light Curve Modeling and Orbital Period Analysis of\\the Contact Binaries OV Leo and V339 Leo}

\author[0009-0007-8508-2357]{Razieh Aliakbari}
\affiliation{Physics Society of Iran, 15875 Tehran, Iran}

\author[0000-0002-0196-9732]{Atila Poro}
\altaffiliation{Corresponding author: atila.poro@obspm.fr}
\affiliation{LUX, Observatoire de Paris, CNRS, PSL, 61 Avenue de l'Observatoire, 75014 Paris, France}

\author[0000-0003-1263-808X]{Raul Michel}
\affiliation{Instituto de Astronom\'ia, UNAM. A.P. 106, 22800 Ensenada, BC, M\'exico}

\author[0009-0004-8426-4114]{Sabrina Baudart}
\affiliation{Double Stars Committee, Société Astronomique de France, 75016 Paris, France}

\begin{abstract}
We present a combined photometric and orbital period analysis of the contact binary systems OV Leo and V339 Leo using ground-based observations together with TESS photometry. The light curves were modeled with the BSN software through a Markov Chain Monte Carlo approach, providing physical estimates of the system parameters and their uncertainties. OV Leo is analyzed photometrically for the first time, while V339 Leo is reexamined using new observations and an updated modeling methodology. The light curve analysis show that both targets are W-subtype contact binaries in shallow-contact configurations. The O'Connell effect detected in V339 Leo is reproduced with a cool starspot model, whereas no spot is required for OV Leo. Three-dimensional geometric models further indicate that OV Leo undergoes total eclipses, while V339 Leo exhibits partial eclipses. Orbital period analysis based on eclipse timing extractions reveals opposite long-term period variations in the two systems, suggesting ongoing mass transfer under the assumption of conservative mass transfer. The absolute parameters were derived using the Gaia parallaxes, the photometric light curve solution, and standard astrophysical equations.
\end{abstract}


\section{Introduction}
In contact binaries, the two components remain enclosed within a common equipotential surface, allowing continuous redistribution of mass and energy, whereas efficient tidal coupling enforces nearly synchronous stellar rotation (\citealt{zahn1977tidal}[1]). Consequently, observable quantities such as the orbital period, mass ratio, and temperature difference reflect the combined effects of mass transfer, angular momentum loss, and energy redistribution within the common envelope, making them important diagnostics of the evolutionary state of contact binaries (\citealt{1968ApJ...151.1123L,1976ApJS...32..583W}[2,3]). This coupling makes contact binaries particularly valuable for testing evolutionary models against observations, because different physical processes can produce measurable changes in their observable properties (\citealt{2026PASP..138c4203S,2024RAA....24e5001P}[4,5]). For example, long-term monotonic period changes are commonly attributed to mass transfer or angular momentum loss, whereas cyclic period variations may arise from magnetic activity through the Applegate mechanism (\citealt{1992ApJ...385..621A}[6]) or from the light-travel time effect (\citealt{1959AJ.....64..149I}[7]) induced by an additional companion. Likewise, the degree of contact and the temperature difference between the components can constrain the efficiency of energy transport through the common envelope, whereas extremely low mass ratios may identify systems approaching tidal instability and possible merger (\citealt{1995ApJ...444L..41R,2006MNRAS.369.2001L,2026ApJ...998..108P}[8,9,10]). Many of these quantities can be inferred through detailed modeling of precise photometric light curves, while spectroscopic observations, when available, provide additional constraints on the system parameters. Applying the same analysis to large samples of contact binaries yields homogeneous sets of physical parameters that can be directly compared with the predictions of binary evolution models. Despite the large number of known contact binaries, accurate light curve solutions combined with reliable orbital period analyses are still unavailable for many systems. Therefore, increasing the sample of well-modeled contact binaries is therefore not simply a matter of improving statistics. It is essential for assessing whether the observed distributions of mass ratio and fillout factor, together with the patterns of orbital period variation, are consistent with current evolutionary models or instead indicate the need for additional physical processes that are not yet fully incorporated into those models.

In this work, we present a combined photometric and orbital period analysis of the contact binaries OV Leo and V339 Leo using Transiting Exoplanet Survey Satellite (TESS; \citealt{2014SPIE.9143E..20R}[11]) and ground-based observations. According to the classifications provided by the All-Sky Automated Survey for SuperNovae (ASAS-SN; \citealt{2018MNRAS.477.3145J}[12]) and the AAVSO Variable Star Index (VSX\footnote{\url{https://vsx.aavso.org/}}), both targets are identified as contact binaries. Basic information for the two binary systems, compiled from Gaia DR3 and the TESS Input Catalog (TIC) v8.2, is presented in Table \ref{Tab:systemsinfo}. According to the VSX database, the orbital periods of OV Leo and V339 Leo are reported as 0.2683061 d and 0.329512 d, respectively, placing both systems among the short-period W UMa-type contact binaries. It should be noted that OV Leo is analyzed photometrically for the first time in the present study. Although V339 Leo was previously investigated by \cite{2025AJ....170..126S}[13], it is reanalyzed using new photometric data together with modern light curve modeling codes and analysis techniques. This work extends the sample of contact binaries investigated within the framework of the Binary Systems of South and North (BSN) project through a homogeneous analysis of both systems.

\begin{table*}
\caption{Coordinates, distance, and temperatures of the system from Gaia DR3 and TIC.}
\centering
\begin{center}
\footnotesize
\begin{tabular}{c c c c c c}
\hline
System & RA$.^\circ$(J2000) & Dec$.^\circ$(J2000) & $d$(pc) & $T_{\text{Gaia}}$(K) & $T_{\text{TIC}}$(K)\\
\hline
OV Leo & 176.837468 & 15.716257 & 470(4) & 5660(9) & 5539(273)\\
V339 Leo & 169.220253 & 14.073507 & 154(1) & 5595(5) & 5815(147)\\
\hline
\end{tabular}
\end{center}
\label{Tab:systemsinfo}
\end{table*}

\vspace{0.4cm}
\section{Observations}
Ground-based multiband photometric observations of OV Leo were obtained at the San Pedro Mártir Observatory, México ($31^\circ02'39''$ N, $115^\circ27'49''$ W; altitude 2830~m), on 2025 March 22. The observations were performed with the 0.84-m $f/15$ Ritchey--Chrétien telescope equipped with a Marconi-5 CCD camera (Spectral Instruments) incorporating an e2v CCD231-42 detector. The detector has a pixel size of $15\times15\,\mu\mathrm{m}^2$, a gain of $2.2~e^-\,\mathrm{ADU}^{-1}$, and a readout noise of $3.6~e^-$. Standard Johnson--Cousins $B$, $V$, $R_{\rm C}$, and $I_{\rm C}$ filters were employed, with exposure times of 90, 50, 35, and 30~s, respectively. The CCD frames were calibrated using standard reduction procedures in IRAF, including bias subtraction and flat-field correction, following the methodology described by \cite{1986SPIE..627..733T}[14]. Differential aperture photometry was carried out using a comparison star at RA = $176.872932^\circ$, Dec = $15.697905^\circ$, and a check star at RA = $176.845608^\circ$, Dec = $15.718040^\circ$ (J2000).

Photometric observations of V339 Leo were carried out at the Observatoire Astronomique de Sabichette (OASa) Observatory in Toulon, France (longitude $05^\circ54'35''$ E, latitude $43^\circ08'59''$ N, altitude 68~m). The system was observed through a standard Johnson $V$ filter on 2024 January 15 and 20 using a 102-mm apochromatic refractor equipped with a ZWO ASI1600MM CCD camera. All images were acquired with $1\times1$ binning and an exposure time of 110~s. Data reduction was performed using three comparison stars and one check star. The equatorial coordinates (J2000) of the comparison stars are: RA = $11^{\rm h}15^{\rm m}48.488^{\rm s}$, Dec = $+14^\circ34'47.85''$; RA = $11^{\rm h}15^{\rm m}05.100^{\rm s}$, Dec = $+14^\circ27'49.0''$; and RA = $11^{\rm h}15^{\rm m}36.061^{\rm s}$, Dec = $+14^\circ36'29.42''$. The check star is located at RA = $11^{\rm h}14^{\rm m}42.342^{\rm s}$ and Dec = $+13^\circ44'50.339''$.

Space-based photometric observations of both binary systems were obtained from the Transiting Exoplanet Survey Satellite (TESS). The satellite surveys the sky with four wide-field cameras, each continuously monitoring an individual sector for approximately 27 days. For each target, we retrieved photometric measurements obtained in the broad TESS passband (600-1000~nm). The corresponding TIC identifiers, the sectors included in the present analysis, and the associated exposure times are summarized in Table~\ref{Tab:tess}. The data were downloaded from the Mikulski Archive for Space Telescopes (MAST) and processed using the Lightkurve Python package, applying detrending procedures aligned with those adopted by the Science Processing Operations Center (SPOC) pipeline before the subsequent analysis.

\begin{table*}
\centering
\footnotesize
\caption{Summary of the TESS observations used in this study. The table lists the TESS sectors, exposure lengths (E.L.), and TIC identifiers for the target systems.}
\begin{tabular}{cc|ccc|c}
\hline
System & TIC & Sector for O-C & Available E.L.(s) & Observation Year & Sector for modeling/E.L.(s)/Pipeline\\
\hline
OV Leo & 	14721942 & 22, 49 & 1800, 600 & 2020, 2022 & 49/600/TESS-SPOC\\
V339 Leo      & 240175453 & 22, 45, 46, 49, 72 & 200, 600, 1800 & 2020, 2021, 2022, 2023 & 72/200/TESS-SPOC\\
\hline
\end{tabular}
\label{Tab:tess}
\end{table*}

\vspace{0.4cm}
\section{Orbital Period Variations}
Possible variations in the orbital periods of the ten target systems were investigated using the O-C analysis by comparing the observed eclipse timings with those calculated from the reference ephemerides (\citealt{2013NewA...21...46L,2019RAA....19..147L,2022AJ....164..202L}[15,16,17]). The times of minima were extracted from several photometric databases and surveys, including ASAS-SN, TESS, VSX, and VarAstro\footnote{\url{https://var.astro.cz/en/}}. The eclipse times from TESS observations were determined by fitting a Gaussian function around the eclipse minima in the light curves. For ASAS-SN and TESS light curves with 1800 s exposure length (\citealt{2019PASP..131f8003B,2019PASP..131a8003M}[18,19]), the data were initially phase-folded using the period-shift method introduced by \cite{2020AJ....159..189L}[20], and the corresponding eclipse timings were subsequently measured. All Heliocentric Julian Dates (HJD) were transformed into Barycentric Julian Dates (BJD) in Barycentric Dynamical Time using \cite{2010PASP..122..935E}[21] method. For VarAstro measurements without reported uncertainties, a fixed error value of $0.001$ was adopted for CCD observations. The eclipse timings obtained from our own photometric observations are listed in Table \ref{Tab:extracted-mins}. The full set of collected minimum timings for the target binaries is provided in the appendix Tables \ref{APP:OVLEO-mins} and \ref{APP:V339LEO-mins}.

\begin{figure*}
\centering
\includegraphics[scale=0.055]{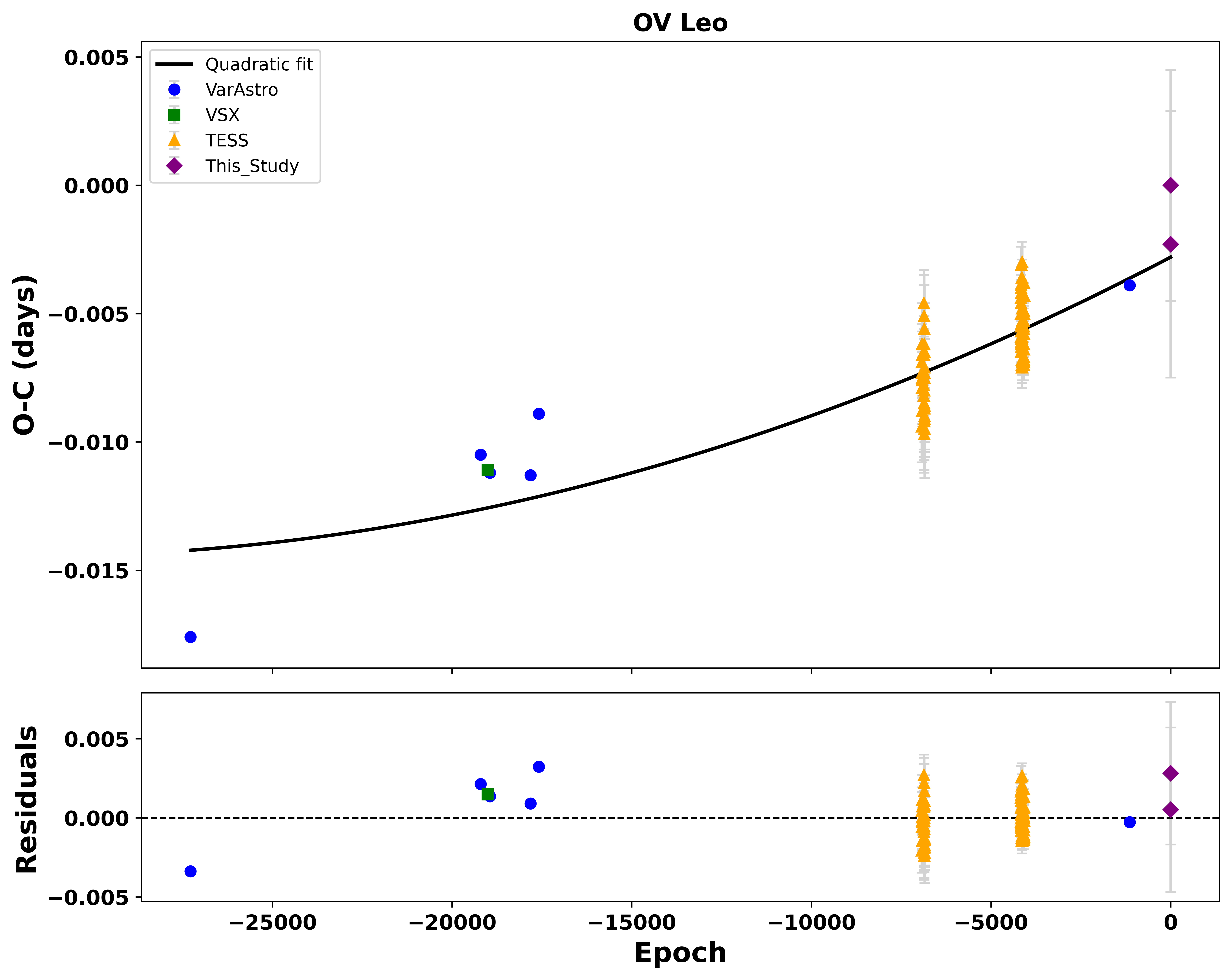}
\includegraphics[scale=0.055]{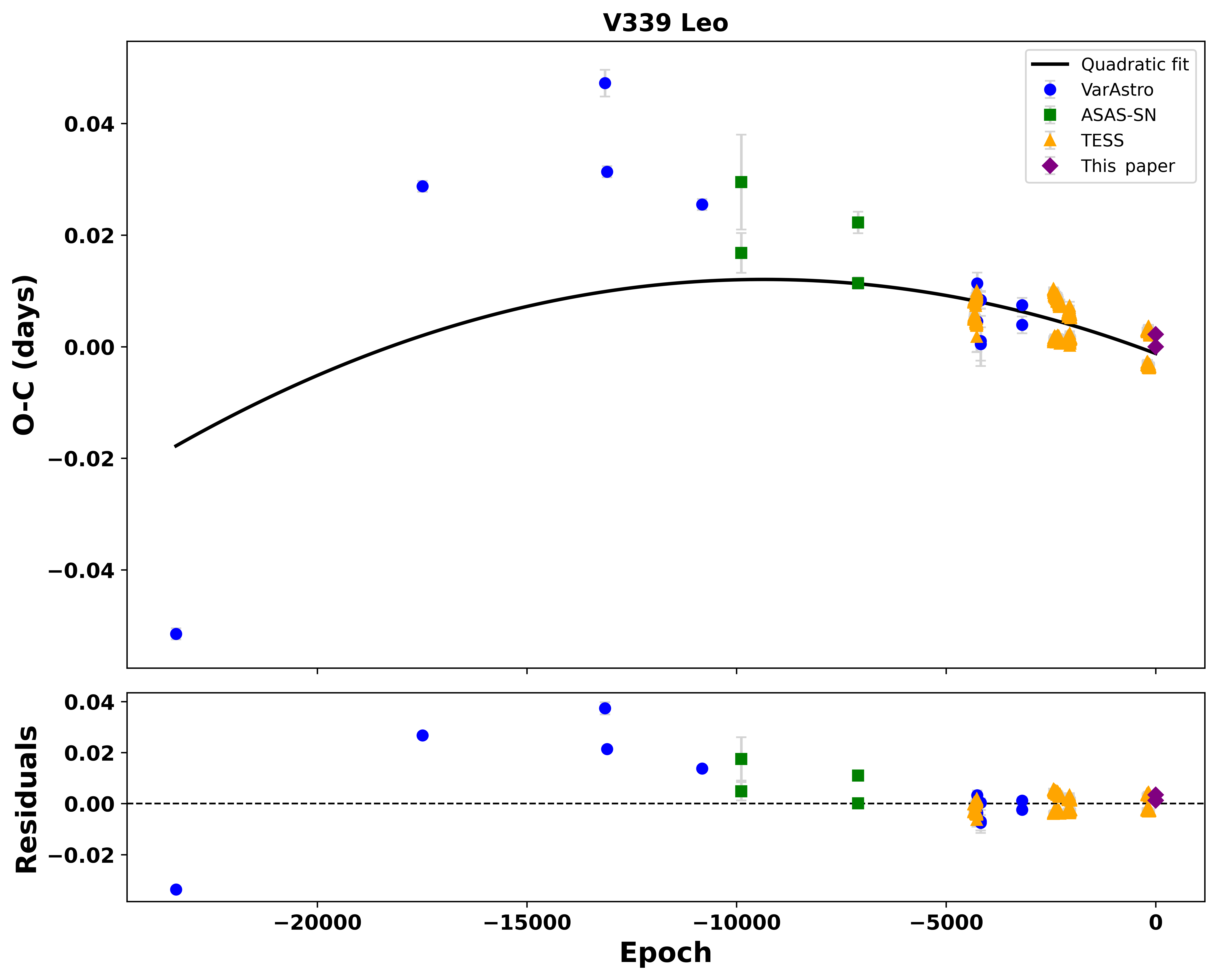}
\caption{$O-C$ diagrams of OV Leo and V339 Leo with the best-fitting models.}
\label{Fig:oc}
\end{figure*}

The O-C values were derived using the linear ephemeris relation:
\begin{equation}
\mathrm{BJD} = \mathrm{BJD}_{0} + P \times E,
\end{equation}
where BJD represents the observed eclipse times, $\mathrm{BJD}_{0}$ (in Table \ref{tab:ephemeris}, it is given as $T_0$) denotes the reference epoch, $P$ is the orbital period, and $E$ corresponds to the orbital cycle number. The resulting O-C values are presented in Tables \ref{Tab:extracted-mins}, \ref{APP:OVLEO-mins}, and \ref{APP:V339LEO-mins}. The O-C diagrams for the target systems are displayed in Figure \ref{Fig:oc}, showing long-term variations in both systems.

For OV Leo and V339 Leo, the O-C variations were modeled using the following quadratic expression:
\begin{equation}
\mathrm{O\!-\!C} = \Delta T_{0} + \Delta P_{0}\times E + \frac{\beta}{2}\times E^{2}.
\end{equation}

The fitting results reveal a positive parabolic trend in the O-C diagram of OV Leo, indicating a continuous increase in its orbital period. In contrast, V339 Leo displays a negative trend, suggesting a long-term decrease in its orbital period. The revised linear ephemerides obtained from these fits are summarized in Table \ref{tab:ephemeris}.

\begin{table*}
\caption{Extracted eclipse minima from the ground-based photometric observations.}
\centering
\small
\begin{tabular}{c c c c c}
\hline
System & Min.($BJD_{TDB}$) & Error  & Epoch  & O-C\\ 
\hline
OV Leo & 2460756.7562 & 0.0045 & 0 & 0\\
       & 2460756.8881 & 0.0052 & 0.5 & -0.0023\\

V339 Leo & 2460330.5782 & 0.0002 & -0.5 & 0.0022 \\
         & 2460330.7407 & 0.0003 & 0	   & 0 \\
\hline
\end{tabular}
\label{Tab:extracted-mins}
\end{table*}

\begin{table*}
\centering
\caption{Reference and revised ephemerides of the studied systems. $T_0$ is the reference epoch of the primary minimum, and $P_0$ is the initial orbital period adopted from the VSX catalog.}
\begin{tabular}{lcccc}
\hline
Target & $T_0$(BJD) &$P$(d) & Corrected $T_0$ & Corrected $P$(d) \\
\hline
OV Leo & 2460756.7562(45) & 0.2683061 & 2460756.7534(5) & 0.2683068(1)\\
V339 Leo & 2460330.7407(3) & 0.329512 & 2460330.7395(3) & 0.3295091(10)\\
\hline
\end{tabular}
\label{tab:ephemeris}
\end{table*}

\vspace{0.4cm}
\section{Light Curve Solution}
The photometric analysis was carried out using the BSN v1.0 software package (\citealt{2025Galax..13...74P}[22]), which was specifically developed for modeling contact binary systems. The modeling combined our ground-based photometric observations with data from a carefully selected TESS sector. To reduce the influence of starspot evolution and other activity-related variations in the light curves, the TESS sector chosen for the analysis was one of the most recent sectors obtained close in time to the ground-based observations (Table~\ref{Tab:tess}). In addition, the selected sector was compared with other recent TESS observations to verify that no significant differences were present.

Before the fitting procedure, a uniform set of physical assumptions was adopted for all target systems. Each binary was analyzed in the contact mode, as indicated by its characteristic light curve morphology, short orbital period, and existing catalog classification, all of which are consistent with a common-envelope configuration in thermal contact. Orbital phases were computed using the ephemerides presented in Table \ref{tab:ephemeris}.

The gravity-darkening coefficients were fixed at $g_{1}=g_{2}=0.32$ (\citealt{1967ZA.....65...89L}[23]), while the bolometric albedos were adopted as $A_{1}=A_{2}=0.5$ (\citealt{1969AcA....19..245R}[24]). The stellar atmospheres were represented using the model of \cite{2004AA...419..725C}[25]. Limb darkening was treated with the logarithmic law implemented in the BSN package, using the coefficients of \cite{1993AJ....106.2096V}[26] together with the updated tables available on the official webpage\footnote{\url{https://faculty.fiu.edu/~vanhamme/limb-darkening/}}, including the coefficients corresponding to the TESS passband.

The initial effective temperatures ($T$) were adopted from the Gaia DR3 catalog (Table~\ref{Tab:systemsinfo}). Following the observed relative depths of the primary and secondary eclipses, the temperatures listed in Gaia DR3 and the TIC were assumed to correspond to the hotter component of each system. The effective temperature of the cooler star was then determined from the observed difference between the eclipse depths.

The light curve of V339 Leo exhibits a clear inequality between its two maxima, indicating the presence of the O'Connell effect. To account for this asymmetry, several spot configurations were explored during the light curve modeling process. Both cool and hot spots were tested on each stellar component, and the resulting solutions were evaluated by comparing their corresponding $\chi^2$ values. The model yielding the lowest $\chi^2$ was adopted as the preferred solution, consistently identifying a unique spot location and temperature. The parameters of the selected model are listed in Table~\ref{tab:analysis}. Surface magnetic activity is generally regarded as the most likely origin of the O'Connell effect in contact binaries, with starspots producing localized brightness variations that distort the observed light curves. Nevertheless, alternative mechanisms have also been proposed to account for this phenomenon (e.g., \citealt{1990ApJ...355..271Z,2003ChJAA...3..142L}[27,28]).

Prior to the final photometric modeling, an initial estimate of the mass ratio was obtained using the MCMC module implemented in the BSN software. Unlike a conventional $q$-search, the MCMC approach allows the parameter space to be explored more efficiently while simultaneously examining the convergence behavior of multiple model parameters. To avoid introducing any prior bias, a wide search interval of $0.05 \leq q \leq 20$ was adopted. For each target system, the MCMC sampling was performed using 28 walkers and 1000 iterations. The entire procedure was repeated three independent times in order to verify the robustness of the inferred solution and to determine whether the optimization converged toward different local minima or consistently approached the same region of parameter space. In both target systems, the three independent runs converged to essentially the same mass ratio value, indicating that the initial estimate is stable and well constrained. During the MCMC search, the effective temperatures of the stellar components were allowed to vary within $\pm100$~K of their adopted initial values. The fillout factor was explored over the interval $0.01 \leq f \leq 1.0$, while the orbital inclination was allowed to vary between $40^\circ$ and $90^\circ$. These parameters were sampled simultaneously with the mass ratio to provide self-consistent initial estimates and to examine whether their convergence was compatible with that obtained for $q$. After the MCMC exploration, the resulting parameter values were refined using the optimization module of the BSN software. The optimized solution obtained at this stage was then adopted as the initial parameter set for the final MCMC analysis presented in the following section.

The final model parameters were inferred from the posterior probability distributions obtained through the MCMC analysis. A total of 28 walkers were evolved for 2000 iterations while sampling the five principal free parameters ($T_1$, $T_2$, $q$, $f$, and $i$). To eliminate the influence of the initial transient phase, the first 400 iterations of each walker were discarded as burn-in. The remaining samples were used to construct the posterior distributions, from which the median was adopted as the final parameter estimate, while the 1$\sigma$ credible intervals were taken as the corresponding uncertainties. The resulting values, listed in Table~\ref{tab:analysis}, provide statistically robust parameter estimates, particularly in cases where the posterior distributions are asymmetric.

\begin{figure*}
\centering
\includegraphics[scale=0.32]{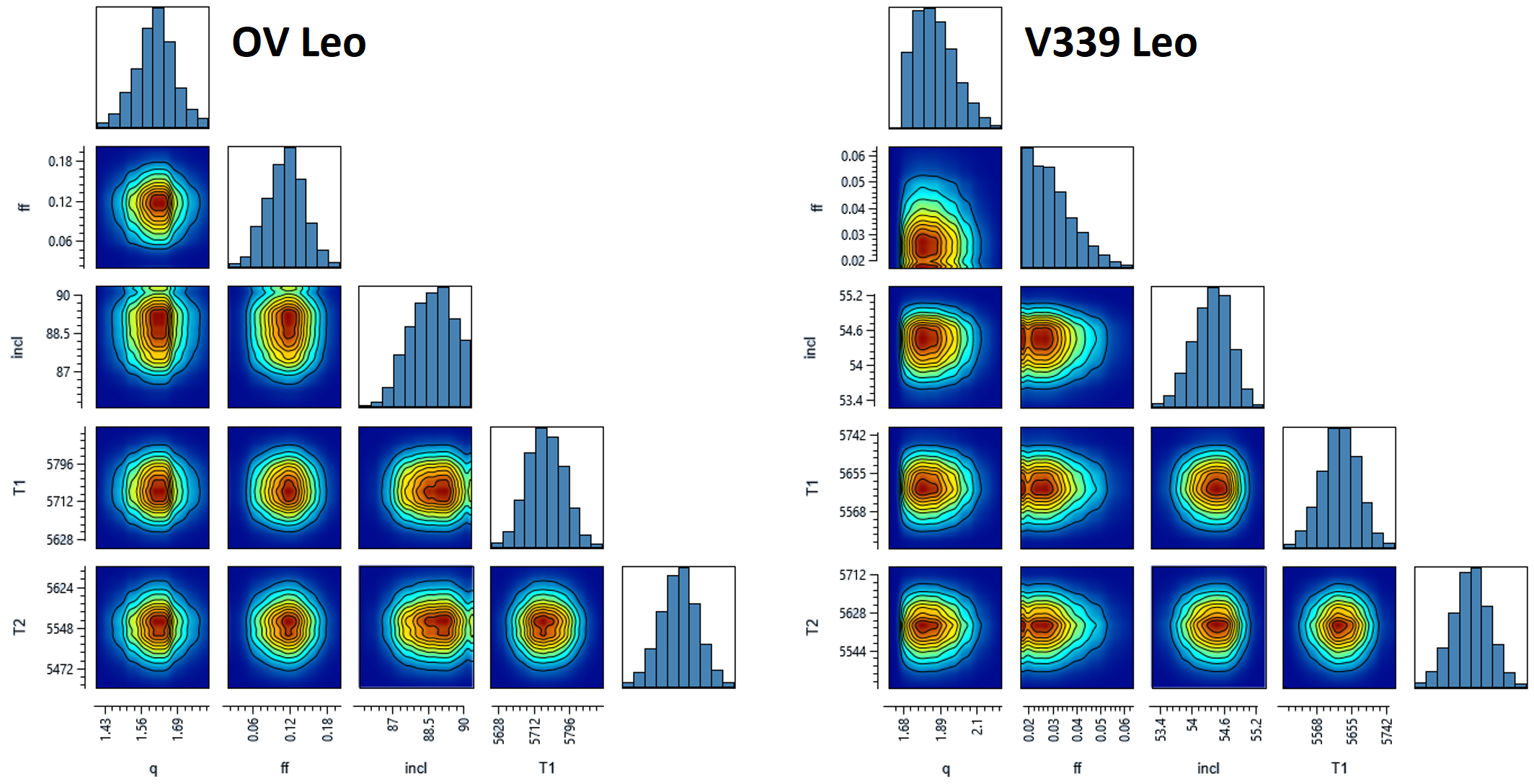}
\caption{Corner plots of the posterior probability distributions for the fitted parameters of two target binaries.}
\label{Fig:corner}
\end{figure*}

\begin{figure*}
\centering
\includegraphics[scale=0.124]{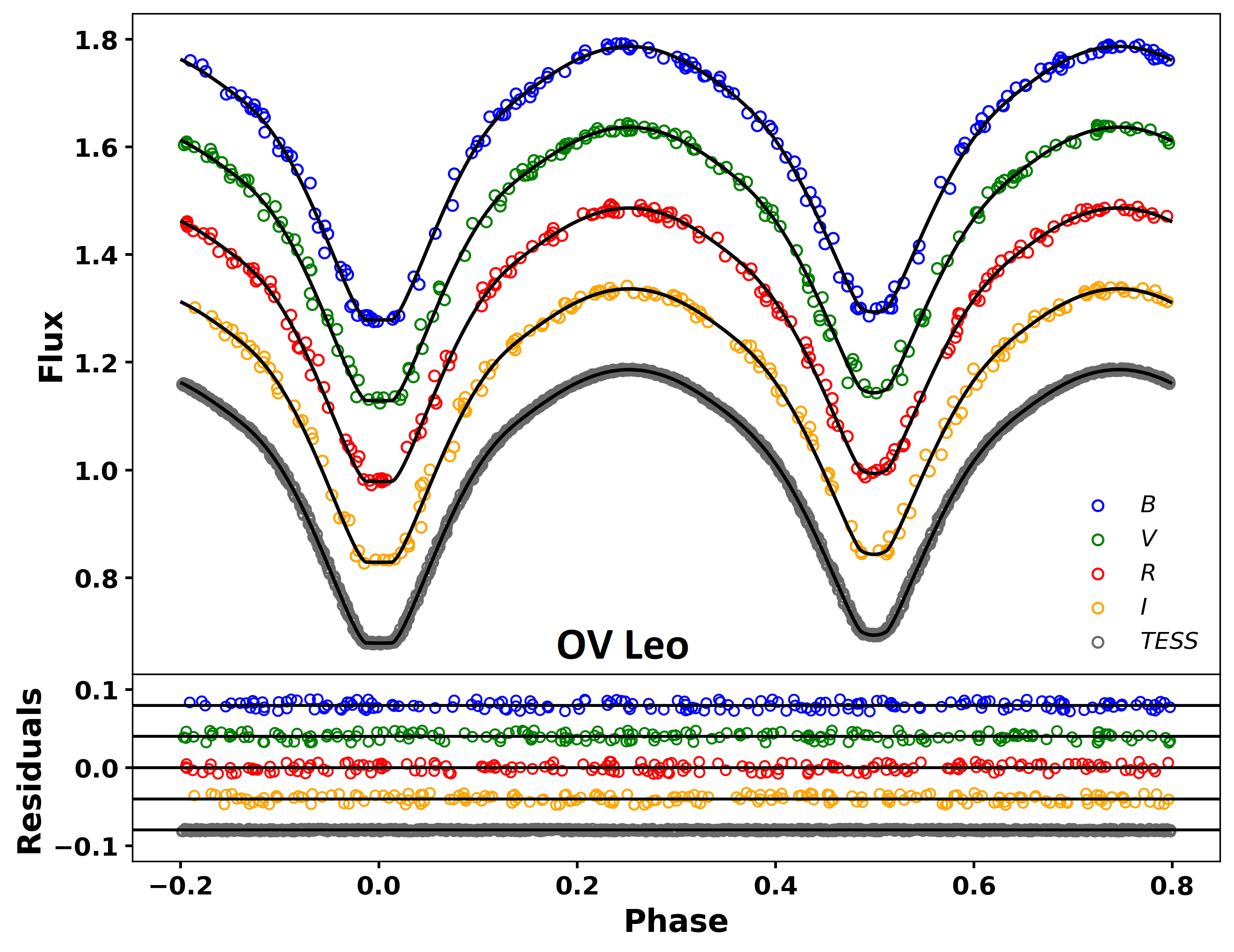}
\includegraphics[scale=0.14]{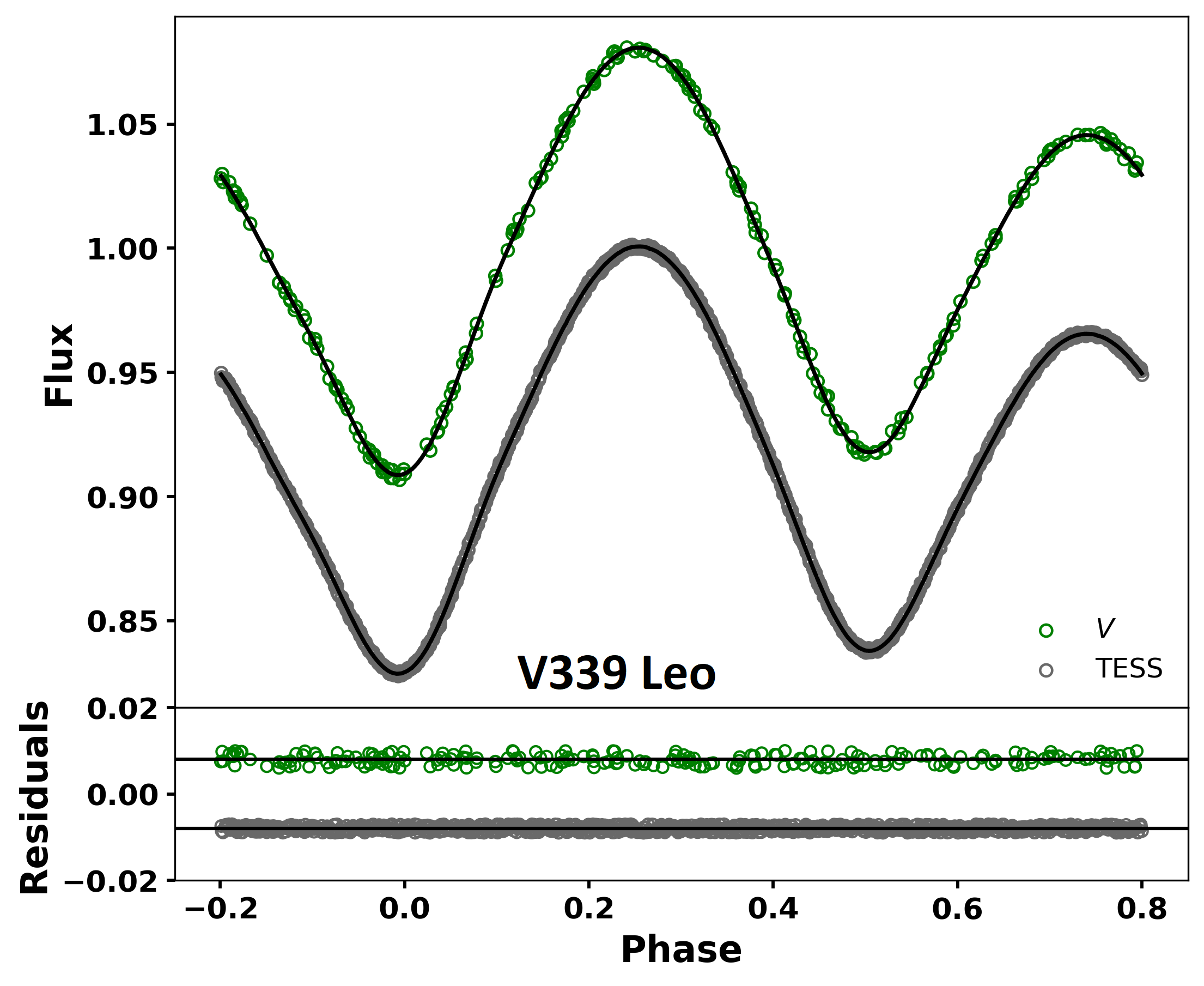}
\caption{Observed and synthetic light curves in the different photometric filters.}
\label{Fig:lc}
\end{figure*}

\begin{figure*}
\centering
\includegraphics[scale=0.36]{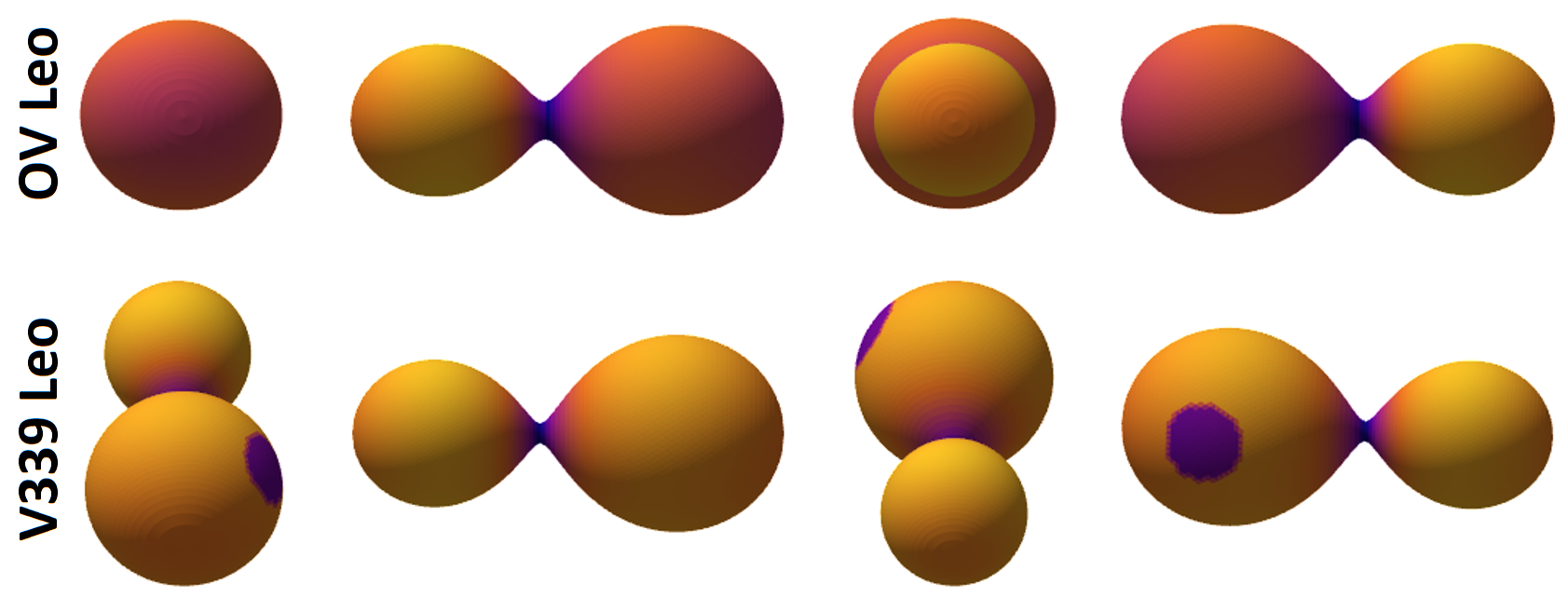}
\caption{Three-dimensional representations of the binary systems at four selected orbital phases (Left to right: 0, 0.25, 0.5, 0.75).}
\label{Fig:3d}
\end{figure*}

The posterior distributions and parameter correlations for OV Leo and V339 Leo are illustrated in the corner diagrams shown in Figure~\ref{Fig:corner}. The corresponding best-fitting synthetic light curves are compared with the observed photometric data in Figure~\ref{Fig:lc}. To provide a visual representation of the final solutions, three-dimensional models of both binary systems are presented in Figure~\ref{Fig:3d}.

\renewcommand\arraystretch{1.4}
\begin{table*}
\centering
\caption{Results of the light curve modeling for the contact binary systems OV Leo and V339 Leo.}
\begin{tabular}{ccc|ccc}
\hline
Parameter &	OV Leo & V339 Leo & Parameter &	OV Leo & V339 Leo\\
\hline					
$T_{1}$ (K) 	&	$5748_{\rm-(41)}^{+(44)}$	&	 $5618_{\rm-(44)}^{+(43)}$ 	&	$r_{(mean)1}$ 	&	0.346(5)	&	 0.328(4)	\\
$T_{2}$ (K) 	&	$5553_{\rm-(39)}^{+(37)}$	&	 $5588_{\rm-(42)}^{+(42)}$ 	&	$r_{(mean)2}$ 	&	0.430(4)	&	0.435(4)	\\
$q=M_2/M_1$ 	&	$1.618_{\rm-(69)}^{+(61)}$	&	 $1.844_{\rm-(104)}^{+(133)}$ 	&		&		&		\\
$i^{\circ}$ 	&	$88.47_{\rm-(99)}^{+(86)}$	&	 $54.34_{\rm-(35)}^{+(30)}$ 	&	$Col.^\circ$(spot) 	&		&	 115(2)	\\
$f$ 	&	$0.114_{\rm-(32)}^{+(29)}$	&	 $0.031_{\rm-(8)}^{+(11)}$ 	&	$Long.^\circ$(spot) 	&		&	 255(2)	\\
$\Omega_1=\Omega_2$ 	&	4.64(5)	&	5.01(3)	&	$Radius^\circ$(spot) 	&		&	 22(1)	\\
$l_1/l_{tot}$ 	&	0.420(5)	&	 0.368(5)	&	$T_{spot}/T_{star}$ 	&		&	 0.89(1)	\\
$l_2/l_{tot}$ 	&	0.580(5)	&	 0.632(5)	&	Component 	&		&	 Secondary	\\
\hline
\end{tabular}
\label{tab:analysis}
\end{table*}

\vspace{0.4cm}
\section{Fundamental Parameters}
The absolute parameters of OV Leo and V339 Leo were derived following the Gaia DR3 parallax-based method described in detail by \cite{2024NewA..11002227P}[29]. This approach combines the geometric information provided by Gaia DR3 with the photometric solution obtained from the light curve analysis, allowing reliable absolute parameters to be estimated in the absence of spectroscopic observations (\citealt{2024RAA....24b5011P}[30]). For both systems, the maximum apparent magnitude ($V_{\rm max}$) was adopted from our own photometric observations (Table~\ref{tab:absolute}). The interstellar extinction in the $V$ band ($A_V$) was taken from the three-dimensional Galactic dust maps of \cite{2019ApJ...887...93G}[31]. These quantities were then used to calculate the absolute visual magnitude of each binary system (Table~\ref{tab:absolute}). The luminosity ratio obtained from the light curve solution was subsequently used to separate the total luminosity into the individual contributions of the two components. Bolometric corrections ($BC$) were adopted from the tables of \cite{1996ApJ...469..355F}[32], and together with the effective temperatures derived from the light curve modeling, the stellar luminosities and radii were calculated through the Stefan-Boltzmann law. The orbital separation was then determined from the fractional radii, and the component masses were obtained by combining the orbital period, semi-major axis, and photometric mass ratio through Kepler's third law. The derived masses, radii, luminosities, bolometric magnitudes, and surface gravities are listed in Table~\ref{tab:absolute}.

\renewcommand\arraystretch{1.2}
\begin{table*}
\centering
\caption{Estimated absolute parameters of the binary components based on the light curve solutions and Gaia DR3 parallax method.}
\begin{tabular}{ccc|ccc}
\hline
Parameter & OV Leo & V339 Leo & Parameter & OV Leo & V339 Leo\\
\hline
$M_1(M_\odot)$ 	&	0.42(7) 	&	0.84(15)	&	$log\textit{(g)}_1$(cgs) 	&	4.48(13)	&	 4.47(14)\\
$M_2(M_\odot)$ 	&	0.67(11)	&	1.55(27)	&	$log\textit{(g)}_2$(cgs) 	&	4.47(13)	&	 4.49(14)\\
$R_1(R_\odot)$ 	&	0.61(4)	&	 0.88(6)	&	$a(R_\odot)$ 	&	1.80(10)	&	 2.68(16)\\
$R_2(R_\odot)$ 	&	0.79(5)	&	 1.17(8)	&		&		&	\\
$L_1(L_\odot)$ 	&	0.36(4)	&	 0.69(7)	&	$V_{max.}$(mag.) & 13.45(10) & 10.15(11)\\				
$L_2(L_\odot)$ 	&	0.53(6)	&	 1.19(13)	&	$A_V$(mag.) & 0.123(1) & 0.044(1)\\				
$M_{bol1}$(mag.) 	&	5.82(11)	&	 5.14(11)	&	$BC_1$(mag.) & -0.085(8) & -0.110(10)\\				
$M_{bol2}$(mag.) 	&	5.43(12)	&	 4.55(12)	&	$BC_2$(mag.) & -0.125(9) & -0.117(9)\\
\hline
\end{tabular}
\label{tab:absolute}
\end{table*}

\vspace{0.4cm}
\section{Dissection and Conclusion}
The light curve analysis of OV Leo is presented for the first time in this study. In contrast, V339 Leo was previously analyzed by \cite{2025AJ....170..126S}[13]. All stages of the light curve analysis for V339 Leo were carried out independently, and the solution was obtained without using the results of the previous study. A detailed comparison between the results obtained in this study for V339 Leo and those reported by \cite{2025AJ....170..126S}[13] highlights agreement in most of the derived system parameters. The orbital inclination derived in the two studies shows very close agreement, with the inclination differing by only $0.5^\circ$. The mass ratios are also in reasonable agreement, with $1/q = 0.419$ reported by \cite{2025AJ....170..126S}[13] and $1/q = 0.542$ obtained in the present study. A pronounced discrepancy, however, appears in the temperature difference between the components: a value of 415~K was reported by \cite{2025AJ....170..126S}[13], whereas the present analysis yields a markedly smaller difference of only 30~K (Table \ref{Tab:conclusion}). The most substantial disagreement concerns the fillout factor, which indicates a shallow contact configuration in our solution, whereas the fillout factor of 28.8\% reported in the previous study corresponds to a medium-contact configuration (Table \ref{Tab:conclusion}). A cool spot is included in the light curve modeling of V339 Leo in both investigations. Differences in methodology and data usage may contribute to these discrepancies. The analysis of \cite{2025AJ....170..126S}[13] is based exclusively on single-band TESS photometry, whereas the present study combines TESS observations with ground-based photometric data obtained in the $V$ band. Interpretations relying solely on single-band photometry, even when derived from space-based observations, should therefore be treated with caution (\citealt{2025MNRAS.538.1427P}[33]). Methodological differences further distinguish the two studies. The work of \cite{2025AJ....170..126S}[13] employs the legacy version of the PHysics Of Eclipsing BinariEs (PHOEBE) code (\citealt{2005ApJ...628..426P}[34]) and does not exploit the iterative capabilities available in more recent implementations. The present analysis applies the BSN application, which incorporates updated physical prescriptions and adopts an MCMC framework for parameter estimation.

The temperature difference between the two components and the spectral classification of each target, adopted from \cite{2018MNRAS.479.5491E}[35], are listed in Table~\ref{Tab:conclusion}. Since the more massive component is not the hotter star in either system, both binaries are classified as W-subtype contact systems (\citealt{1970VA.....12..217B}[36]). Furthermore, according to the classification scheme proposed by \cite{2022AJ....164..202L}[17], both systems are classified as shallow-contact binaries based on our light curve analysis. A total eclipse occurs when one component is completely obscured by its companion during eclipse (\citealt{2025PASP..137l4202P}[37]). Based on the three-dimensional models shown in Figure~\ref{Fig:3d}, OV Leo is therefore classified as a totally eclipsing system, whereas V339 Leo exhibits partial eclipses.

The parabolic trends in the O-C diagrams indicate opposite orbital period variations for the two systems, with an increasing orbital period for OV Leo and a decreasing orbital period for V339 Leo. These long-term orbital period variations may be attributed to mass transfer between the binary components (\citealt{1998CoSka..28..101P}[38]). Assuming conservative mass transfer, the corresponding mass-transfer rates were calculated from the observed orbital period variations using
\begin{equation}
\dot{M}_1=\frac{\dot{P}\,M_1M_2}{3P\left(M_1-M_2\right)}.
\end{equation}
where $\dot{M}_1$ is the mass-transfer rate of the primary component, $\dot{P}$ is the orbital period change rate, $M_1$ and $M_2$ are the masses of the primary and secondary components, respectively, and $P$ is the orbital period. The resulting mass-transfer rates are listed in Table~\ref{Tab:conclusion}.

\begin{table*}
\caption{Physical and evolutionary properties of the studied contact binaries, including temperature differences, spectral classifications, contact degrees, mass-transfer rates, and orbital period changes.}
\centering
\begin{center}
\footnotesize
\begin{tabular}{ccccc|cc}
\hline
System & $|\Delta T|$ (K) & Sp. type & Subtype & $f$ classification & $\dot{M}_1$ ($10^{-7}\,M_{\odot}\,\mathrm{yr}^{-1}$) & $\dot{P}$ ($10^{-8}\,\mathrm{d\,yr^{-1}}$)\\
\hline
OV Leo & 195 & G5-G8 & W & Shallow & -0.44(36) & 3.14(1.23)\\
V339 Leo & 30 & G7-G8 & W & Shallow & 6.25(3.74) & -3.37(21)\\
\hline
\end{tabular}
\end{center}
\label{Tab:conclusion}
\end{table*}

\vspace{0.4cm}
\section*{Data Availability}
The ground-based observations used in this study are available from the corresponding author upon reasonable request. The eclipse timings extracted from different TESS sectors for V339 Leo are numerous; therefore, only a subset of these measurements is provided in the supplementary table. The complete set of extracted eclipse timings is available from the corresponding author upon reasonable request.

\vspace{0.4cm}
\section*{Acknowledgments}
This manuscript, including the observation, analysis, and writing processes, was provided by the BSN project (\url{https://bsnp.info}). Photometric analysis of OV Leo is based on observations carried out at the Observatorio Astron\'omico Nacional on the Sierra San Pedro M\'artir, operated by the Universidad Nacional Aut\'onoma de M\'exico. Data reduction was performed using IRAF, distributed by the National Optical Observatories and operated by the Association of Universities for Research in Astronomy, Inc., under a cooperative agreement with the National Science Foundation. This study also makes use of results from the European Space Agency's Gaia mission (\url{http://www.cosmos.esa.int/gaia}) and of observations obtained by the TESS mission, supported through NASA's Explorer Program. We are grateful to Kai Li for valuable discussions and for his helpful insights into the eclipse timing analysis.

\vspace{0.4cm}
\section*{Declarations}
Funding: No funds, grants, or other support was received.

Competing interests: The authors have no competing interests to declare that are relevant to the content of this paper.

\section*{Appendix}
The continuation of Table~\ref{Tab:extracted-mins}, which lists the extracted eclipse minimum times from TESS photometry and the literature for the two target systems, is provided in the Appendix.

\begin{table*}
\renewcommand\arraystretch{0.9}
\caption{Collected and extracted times of minima for OV Leo.}
\centering
\small
\begin{tabular}{cccccccccc}
\hline
Min. & Error  & Epoch  & O-C & Ref. & Min. & Error  & Epoch  & O-C & Ref.\\ 
\hline
2453438.6897	&		&	-27275.0	&	-0.0176	&	VarAstro	&	2459640.3288	&	0.0002	&	-4161.0	&	-0.0057	&	TESS	\\
2455604.8661	&		&	-19201.5	&	-0.0105	&	VarAstro	&	2459640.4628	&	0.0004	&	-4160.5	&	-0.0059	&	TESS	\\
2455656.7828	&		&	-19008.0	&	-0.0111	&	VSX	&	2459640.5989	&	0.0006	&	-4160.0	&	-0.0039	&	TESS	\\
2455674.7592	&		&	-18941.0	&	-0.0112	&	VarAstro	&	2459640.7320	&	0.0002	&	-4159.5	&	-0.0050	&	TESS	\\
2455976.8717	&		&	-17815.0	&	-0.0113	&	VarAstro	&	2459640.8652	&	0.0002	&	-4159.0	&	-0.0060	&	TESS	\\
2456038.7187	&		&	-17584.5	&	-0.0089	&	VarAstro	&	2459641.0013	&	0.0005	&	-4158.5	&	-0.0040	&	TESS	\\
2458899.4003	&	0.0015	&	-6922.5	&	-0.0069	&	TESS	&	2459641.1332	&	0.0002	&	-4158.0	&	-0.0063	&	TESS	\\
2458899.8003	&	0.0014	&	-6921.0	&	-0.0094	&	TESS	&	2459641.2676	&	0.0003	&	-4157.5	&	-0.0060	&	TESS	\\
2458899.9360	&	0.0011	&	-6920.5	&	-0.0079	&	TESS	&	2459641.4036	&	0.0005	&	-4157.0	&	-0.0042	&	TESS	\\
2458899.9360	&	0.0011	&	-6920.5	&	-0.0079	&	TESS	&	2459641.5365	&	0.0002	&	-4156.5	&	-0.0054	&	TESS	\\
2458900.2043	&	0.0003	&	-6919.5	&	-0.0079	&	TESS	&	2459641.6699	&	0.0002	&	-4156.0	&	-0.0061	&	TESS	\\
2458900.3387	&	0.0019	&	-6919.0	&	-0.0076	&	TESS	&	2459641.8071	&	0.0007	&	-4155.5	&	-0.0031	&	TESS	\\
2458900.4743	&	0.0016	&	-6918.5	&	-0.0062	&	TESS	&	2459645.9621	&	0.0005	&	-4140.0	&	-0.0069	&	TESS	\\
2458900.6073	&	0.0009	&	-6918.0	&	-0.0073	&	TESS	&	2459646.0995	&	0.0007	&	-4139.5	&	-0.0036	&	TESS	\\
2458901.1424	&	0.0006	&	-6916.0	&	-0.0088	&	TESS	&	2459646.2311	&	0.0002	&	-4139.0	&	-0.0062	&	TESS	\\
2458901.4120	&	0.0008	&	-6915.0	&	-0.0075	&	TESS	&	2459646.3644	&	0.0007	&	-4138.5	&	-0.0070	&	TESS	\\
2458901.5471	&	0.0012	&	-6914.5	&	-0.0066	&	TESS	&	2459646.5018	&	0.0006	&	-4138.0	&	-0.0038	&	TESS	\\
2458902.0823	&	0.0014	&	-6912.5	&	-0.0079	&	TESS	&	2459646.6344	&	0.0002	&	-4137.5	&	-0.0053	&	TESS	\\
2458914.2896	&	0.0009	&	-6867.0	&	-0.0087	&	TESS	&	2459646.7671	&	0.0004	&	-4137.0	&	-0.0068	&	TESS	\\
2458914.5583	&	0.0001	&	-6866.0	&	-0.0082	&	TESS	&	2459646.9009	&	0.0008	&	-4136.5	&	-0.0071	&	TESS	\\
2458914.8302	&	0.0013	&	-6865.0	&	-0.0046	&	TESS	&	2459647.0372	&	0.0003	&	-4136.0	&	-0.0050	&	TESS	\\
2458914.9624	&	0.0013	&	-6864.5	&	-0.0066	&	TESS	&	2459647.0372	&	0.0003	&	-4136.0	&	-0.0050	&	TESS	\\
2458915.2295	&	0.0012	&	-6863.5	&	-0.0078	&	TESS	&	2459647.1704	&	0.0003	&	-4135.5	&	-0.0059	&	TESS	\\
2458916.0371	&	0.0016	&	-6860.5	&	-0.0051	&	TESS	&	2459647.3034	&	0.0005	&	-4135.0	&	-0.0071	&	TESS	\\
2458916.7044	&	0.0002	&	-6858.0	&	-0.0085	&	TESS	&	2459647.4398	&	0.0002	&	-4134.5	&	-0.0048	&	TESS	\\
2458916.9751	&	0.0014	&	-6857.0	&	-0.0062	&	TESS	&	2459647.5724	&	0.0003	&	-4134.0	&	-0.0064	&	TESS	\\
2458917.1090	&	0.0014	&	-6856.5	&	-0.0065	&	TESS	&	2459647.7099	&	0.0008	&	-4133.5	&	-0.0030	&	TESS	\\
2458917.3740	&	0.0015	&	-6855.5	&	-0.0097	&	TESS	&	2459647.8421	&	0.0003	&	-4133.0	&	-0.0050	&	TESS	\\
2458917.5084	&	0.0016	&	-6855.0	&	-0.0095	&	TESS	&	2459647.9752	&	0.0002	&	-4132.5	&	-0.0060	&	TESS	\\
2458917.6441	&	0.0009	&	-6854.5	&	-0.0080	&	TESS	&	2459648.1086	&	0.0004	&	-4132.0	&	-0.0068	&	TESS	\\
2458917.9128	&	0.0003	&	-6853.5	&	-0.0075	&	TESS	&	2459648.2453	&	0.0005	&	-4131.5	&	-0.0042	&	TESS	\\
2458918.0472	&	0.0017	&	-6853.0	&	-0.0073	&	TESS	&	2459648.3783	&	0.0002	&	-4131.0	&	-0.0054	&	TESS	\\
2458918.1831	&	0.0017	&	-6852.5	&	-0.0056	&	TESS	&	2459660.5862	&	0.0002	&	-4085.5	&	-0.0055	&	TESS	\\
2458918.3136	&	0.0015	&	-6852.0	&	-0.0092	&	TESS	&	2459660.7189	&	0.0005	&	-4085.0	&	-0.0069	&	TESS	\\
2458918.5821	&	0.0016	&	-6851.0	&	-0.0090	&	TESS	&	2459660.9885	&	0.0001	&	-4084.0	&	-0.0056	&	TESS	\\
2458918.7166	&	0.0018	&	-6850.5	&	-0.0086	&	TESS	&	2459661.1212	&	0.0006	&	-4083.5	&	-0.0070	&	TESS	\\
2458918.8508	&	0.0004	&	-6850.0	&	-0.0087	&	TESS	&	2459661.2586	&	0.0006	&	-4083.0	&	-0.0038	&	TESS	\\
2458919.1206	&	0.0011	&	-6849.0	&	-0.0071	&	TESS	&	2459661.3913	&	0.0001	&	-4082.5	&	-0.0053	&	TESS	\\
2458919.2554	&	0.0019	&	-6848.5	&	-0.0065	&	TESS	&	2459661.5240	&	0.0004	&	-4082.0	&	-0.0067	&	TESS	\\
2458919.2554	&	0.0019	&	-6848.5	&	-0.0065	&	TESS	&	2459661.7940	&	0.0003	&	-4081.0	&	-0.0050	&	TESS	\\
2458919.6552	&	0.0012	&	-6847.0	&	-0.0091	&	TESS	&	2459661.9269	&	0.0003	&	-4080.5	&	-0.0062	&	TESS	\\
2458920.7281	&	0.0019	&	-6843.0	&	-0.0095	&	TESS	&	2459662.0605	&	0.0005	&	-4080.0	&	-0.0068	&	TESS	\\
2458920.8631	&	0.0014	&	-6842.5	&	-0.0086	&	TESS	&	2459662.1966	&	0.0003	&	-4079.5	&	-0.0049	&	TESS	\\
2459639.3910	&	0.0003	&	-4164.5	&	-0.0044	&	TESS	&	2459662.3292	&	0.0003	&	-4079.0	&	-0.0064	&	TESS	\\
2459639.5234	&	0.0002	&	-4164.0	&	-0.0062	&	TESS	&	2459662.5989	&	0.0003	&	-4078.0	&	-0.0050	&	TESS	\\
2459639.6576	&	0.0005	&	-4163.5	&	-0.0062	&	TESS	&	2459662.7322	&	0.0002	&	-4077.5	&	-0.0058	&	TESS	\\
2459639.7933	&	0.0004	&	-4163.0	&	-0.0046	&	TESS	&	2459662.8654	&	0.0004	&	-4077.0	&	-0.0069	&	TESS	\\
2459639.9265	&	0.0002	&	-4162.5	&	-0.0055	&	TESS	&	2459663.0021	&	0.0005	&	-4076.5	&	-0.0043	&	TESS	\\
2459640.0597	&	0.0004	&	-4162.0	&	-0.0065	&	TESS	&	2460451.4200	&		&	-1138.0	&	-0.0039	&	VarAstro	\\
2459640.1944	&	0.0003	&	-4161.5	&	-0.0060	&	TESS	&		&		&		&		&		\\
\hline
\end{tabular}
\label{APP:OVLEO-mins}
\end{table*}

\begin{table*}
\renewcommand\arraystretch{0.9}
\caption{Collected and extracted times of minima for V339 Leo.}
\centering
\small
\begin{tabular}{cccccccccc}
\hline
Min. & Error  & Epoch  & O-C & Ref. & Min. & Error  & Epoch  & O-C & Ref.\\ 
\hline
2452629.1700	&	0.0010	&	-23372.5	&	-0.0515	&	VarAstro	&	2459552.9363	&	0.0002	&	-2360.5	&	0.0087	&	TESS	\\
2454566.6160	&	0.0010	&	-17493.0	&	0.0288	&	VarAstro	&	2459553.0938	&	0.0003	&	-2360.0	&	0.0014	&	TESS	\\
2456001.4945	&	0.0024	&	-13138.5	&	0.0473	&	VarAstro	&	2459553.2658	&	0.0003	&	-2359.5	&	0.0087	&	TESS	\\
2456018.4485	&	0.0010	&	-13087.0	&	0.0314	&	VarAstro	&	2459553.4234	&	0.0003	&	-2359.0	&	0.0015	&	TESS	\\
2456765.4463	&	0.0010	&	-10820.0	&	0.0255	&	VarAstro	&	2459553.5952	&	0.0002	&	-2358.5	&	0.0086	&	TESS	\\
2457072.5555	&	0.0085	&	-9888.0	&	0.0295	&	ASAS-SN	&	2459553.7528	&	0.0003	&	-2358.0	&	0.0015	&	TESS	\\
2457072.7076	&	0.0036	&	-9887.5	&	0.0168	&	ASAS-SN	&	2459553.9245	&	0.0002	&	-2357.5	&	0.0083	&	TESS	\\
2457990.7335	&	0.0019	&	-7101.5	&	0.0223	&	ASAS-SN	&	2459554.0825	&	0.0003	&	-2357.0	&	0.0016	&	TESS	\\
2457990.8874	&	0.0010	&	-7101.0	&	0.0114	&	ASAS-SN	&	2459571.0586	&	0.0002	&	-2305.5	&	0.0078	&	TESS	\\
2458899.5103	&	0.0002	&	-4343.5	&	0.0049	&	TESS	&	2459571.2172	&	0.0003	&	-2305.0	&	0.0017	&	TESS	\\
2458899.6782	&	0.0003	&	-4343.0	&	0.0082	&	TESS	&	2459571.3874	&	0.0002	&	-2304.5	&	0.0071	&	TESS	\\
2458899.8397	&	0.0002	&	-4342.5	&	0.0049	&	TESS	&	2459571.5467	&	0.0003	&	-2304.0	&	0.0017	&	TESS	\\
2458900.0076	&	0.0003	&	-4342.0	&	0.0080	&	TESS	&	2459571.7174	&	0.0002	&	-2303.5	&	0.0076	&	TESS	\\
2458900.1694	&	0.0003	&	-4341.5	&	0.0051	&	TESS	&	2459571.8759	&	0.0003	&	-2303.0	&	0.0014	&	TESS	\\
2458900.3371	&	0.0002	&	-4341.0	&	0.0080	&	TESS	&	2459572.0469	&	0.0002	&	-2302.5	&	0.0076	&	TESS	\\
2458900.4988	&	0.0002	&	-4340.5	&	0.0050	&	TESS	&	2459572.2054	&	0.0003	&	-2302.0	&	0.0013	&	TESS	\\
2458900.6665	&	0.0002	&	-4340.0	&	0.0079	&	TESS	&	2459572.3768	&	0.0003	&	-2301.5	&	0.0080	&	TESS	\\
2458900.8285	&	0.0002	&	-4339.5	&	0.0052	&	TESS	&	2459572.5346	&	0.0003	&	-2301.0	&	0.0010	&	TESS	\\
2458900.9961	&	0.0002	&	-4339.0	&	0.0079	&	TESS	&	2459647.9927	&	0.0003	&	-2072.0	&	0.0009	&	TESS	\\
2458916.9743	&	0.0005	&	-4290.5	&	0.0048	&	TESS	&	2459648.1624	&	0.0003	&	-2071.5	&	0.0059	&	TESS	\\
2458917.1430	&	0.0003	&	-4290.0	&	0.0088	&	TESS	&	2459648.3223	&	0.0003	&	-2071.0	&	0.0010	&	TESS	\\
2458917.3035	&	0.0002	&	-4289.5	&	0.0046	&	TESS	&	2459648.4924	&	0.0002	&	-2070.5	&	0.0063	&	TESS	\\
2458917.4725	&	0.0003	&	-4289.0	&	0.0087	&	TESS	&	2459648.6518	&	0.0003	&	-2070.0	&	0.0010	&	TESS	\\
2458917.6328	&	0.0003	&	-4288.5	&	0.0044	&	TESS	&	2459648.8223	&	0.0003	&	-2069.5	&	0.0067	&	TESS	\\
2458917.8021	&	0.0003	&	-4288.0	&	0.0089	&	TESS	&	2459648.9813	&	0.0003	&	-2069.0	&	0.0010	&	TESS	\\
2458917.9621	&	0.0003	&	-4287.5	&	0.0041	&	TESS	&	2459649.1517	&	0.0002	&	-2068.5	&	0.0066	&	TESS	\\
2458918.1316	&	0.0003	&	-4287.0	&	0.0089	&	TESS	&	2459649.3111	&	0.0003	&	-2068.0	&	0.0013	&	TESS	\\
2458918.2916	&	0.0003	&	-4286.5	&	0.0041	&	TESS	&	2459649.4813	&	0.0003	&	-2067.5	&	0.0066	&	TESS	\\
2458918.4611	&	0.0003	&	-4286.0	&	0.0089	&	TESS	&	2460263.3581	&	0.0003	&	-204.5	&	0.0026	&	TESS	\\
2458927.3604	&	0.0019	&	-4259.0	&	0.0114	&	VarAstro	&	2460263.5172	&	0.0004	&	-204.0	&	-0.0031	&	TESS	\\
2458927.5183	&	0.0054	&	-4258.5	&	0.0045	&	VarAstro	&	2460263.6879	&	0.0003	&	-203.5	&	0.0029	&	TESS	\\
2458954.3774	&	0.0016	&	-4177.0	&	0.0084	&	VarAstro	&	2460263.8466	&	0.0004	&	-203.0	&	-0.0032	&	TESS	\\
2458954.5348	&	0.0045	&	-4176.5	&	0.0010	&	VarAstro	&	2460264.0174	&	0.0003	&	-202.5	&	0.0029	&	TESS	\\
2458955.3659	&	0.0015	&	-4174.0	&	0.0084	&	VarAstro	&	2460264.1763	&	0.0004	&	-202.0	&	-0.0030	&	TESS	\\
2458955.5228	&	0.0030	&	-4173.5	&	0.0005	&	VarAstro	&	2460264.3470	&	0.0003	&	-201.5	&	0.0030	&	TESS	\\
2459280.4287	&	0.0013	&	-3187.5	&	0.0075	&	VarAstro	&	2460264.5057	&	0.0004	&	-201.0	&	-0.0031	&	TESS	\\
2459280.5899	&	0.0015	&	-3187.0	&	0.0039	&	VarAstro	&	2460264.6765	&	0.0003	&	-200.5	&	0.0029	&	TESS	\\
2459532.9931	&	0.0003	&	-2421.0	&	0.0010	&	TESS	&	2460264.8351	&	0.0004	&	-200.0	&	-0.0032	&	TESS	\\
2459533.1659	&	0.0002	&	-2420.5	&	0.0090	&	TESS	&	2460280.6511	&	0.0004	&	-152.0	&	-0.0038	&	TESS	\\
2459533.3232	&	0.0003	&	-2420.0	&	0.0016	&	TESS	&	2460280.8222	&	0.0003	&	-151.5	&	0.0026	&	TESS	\\
2459533.4957	&	0.0002	&	-2419.5	&	0.0093	&	TESS	&	2460280.9809	&	0.0004	&	-151.0	&	-0.0035	&	TESS	\\
2459533.6525	&	0.0003	&	-2419.0	&	0.0013	&	TESS	&	2460281.1518	&	0.0003	&	-150.5	&	0.0027	&	TESS	\\
2459533.8252	&	0.0002	&	-2418.5	&	0.0093	&	TESS	&	2460281.3103	&	0.0004	&	-150.0	&	-0.0036	&	TESS	\\
2459533.9822	&	0.0003	&	-2418.0	&	0.0015	&	TESS	&	2460281.4813	&	0.0003	&	-149.5	&	0.0027	&	TESS	\\
2459534.1547	&	0.0002	&	-2417.5	&	0.0092	&	TESS	&	2460281.6396	&	0.0004	&	-149.0	&	-0.0038	&	TESS	\\
2459534.3114	&	0.0003	&	-2417.0	&	0.0012	&	TESS	&	2460281.8107	&	0.0003	&	-148.5	&	0.0025	&	TESS	\\
2459534.4849	&	0.0003	&	-2416.5	&	0.0100	&	TESS	&	2460281.9692	&	0.0004	&	-148.0	&	-0.0037	&	TESS	\\
2459551.7754	&	0.0003	&	-2364.0	&	0.0010	&	TESS	&	2460282.1400	&	0.0003	&	-147.5	&	0.0024	&	TESS	\\
2459552.7638	&	0.0003	&	-2361.0	&	0.0010	&	TESS	&	2460285.4351	&	0.0003	&	-137.5	&	0.0023	&	TESS	\\
\hline
\end{tabular}
\label{APP:V339LEO-mins}
\end{table*}


\end{document}